\documentclass[twocolumn,english,aps,prl,reprint,superscriptaddress]{revtex4-1}
\usepackage[T1]{fontenc}
\usepackage[latin9]{inputenc}
\usepackage{amsmath}
\usepackage{amssymb}
\usepackage{graphicx}
\usepackage{subscript}
\usepackage{wasysym}

\makeatletter
\@ifundefined{textcolor}{}
{%
 \definecolor{BLACK}{gray}{0}
 \definecolor{WHITE}{gray}{1}
 \definecolor{RED}{rgb}{1,0,0}
 \definecolor{GREEN}{rgb}{0,1,0}
 \definecolor{BLUE}{rgb}{0,0,1}
 \definecolor{CYAN}{cmyk}{1,0,0,0}
 \definecolor{MAGENTA}{cmyk}{0,1,0,0}
 \definecolor{YELLOW}{cmyk}{0,0,1,0}
}

\usepackage{pslatex}

\usepackage{babel}

\usepackage{babel}

\usepackage{babel}

\makeatother

\usepackage{babel}
\begin{document}

\title{Coherent Bichromatic Force Deflection of Molecules}

\author{Ivan Kozyryev}

\thanks{These authors contributed equally. ivan@cua.harvard.edu, louis@cua.harvard.edu}


\affiliation{Harvard-MIT Center for Ultracold Atoms, Cambridge, MA 02138, USA}

\affiliation{Department of Physics, Harvard University, Cambridge, MA 02138, USA}



\author{Louis Baum}

\thanks{These authors contributed equally. ivan@cua.harvard.edu, louis@cua.harvard.edu}


\affiliation{Harvard-MIT Center for Ultracold Atoms, Cambridge, MA 02138, USA}

\affiliation{Department of Physics, Harvard University, Cambridge, MA 02138, USA}

\author{Leland Aldridge}

\altaffiliation[Present address: ]{Department of Physics, Yale University, New Haven, CT 06520, USA}

\affiliation{Department of Physics, University of Connecticut, Storrs, CT 06269, USA}

\author{Phelan Yu}

\affiliation{Harvard-MIT Center for Ultracold Atoms, Cambridge, MA 02138, USA}

\affiliation{Department of Physics, Harvard University, Cambridge, MA 02138, USA}

\author{Edward E. Eyler}

\affiliation{Department of Physics, University of Connecticut, Storrs, CT 06269, USA}

\author{John M. Doyle}

\affiliation{Harvard-MIT Center for Ultracold Atoms, Cambridge, MA 02138, USA}

\affiliation{Department of Physics, Harvard University, Cambridge, MA 02138, USA}

\date{\today}

\begin{abstract}
We demonstrate the coherent optical bichromatic force on a molecule, the polar free radical strontium monohydroxide (SrOH). A dual-frequency retro-reflected laser beam addressing the $\tilde{X}^2\Sigma^+\leftrightarrow\tilde{A}^2\Pi_{1/2}$ electronic transition  coherently imparts momentum onto a cryogenic beam of SrOH. This directional photon exchange creates  a bichromatic force that transversely deflects the molecules.  By adjusting the relative phase between the forward and counter propagating laser beams we reverse the direction of the applied force.
A momentum transfer of $70\hbar k$ is achieved with minimal loss of molecules to dark states. 
Modeling of the bichromatic force is performed via direct numerical solution of the time-dependent density matrix and is compared with experimental observations. Our results open the door  to further coherent manipulation of molecular motion, including the efficient optical deceleration of diatomic and polyatomic molecules with complex level structures. 

\end{abstract}
\maketitle


%
Laser manipulation of atomic motion has revolutionized atomic, molecular and optical (AMO) physics \cite{chu1998nobel, phillips1998nobel}.
The widely-used techniques of laser cooling and trapping made possible the creation of ultracold degenerate quantum gases \cite{ketterle2002nobel}, simulation of important condensed matter models \cite{bloch2008many} and development of new quantum sensors \cite{dickerson2013multiaxis, cronin2009optics} and clocks \cite{ludlow2015optical}. Laser deceleration and cooling of atomic beams -- a necessary part of the trap loading process -- typically requires
scattering tens of thousands of photons in order to bring room (or oven) temperature
atoms to velocities where they can be confined
by electromagnetic traps for further studies \cite{metcalf2012laser}. While beam deceleration employing the spontaneous radiation pressure force has been
a standard for atomic experiments, its application
to slowing molecular beams has been limited by the small change
in kinetic energy per scattered photon and the myriad of internal molecular states, which inhibits photon cycling. At the same time, there is extreme interest in creating ultracold molecules for new physics applications \cite{Carr2009}.    

\begin{figure}[h]
\begin{centering}
\includegraphics[width=0.4\textwidth]{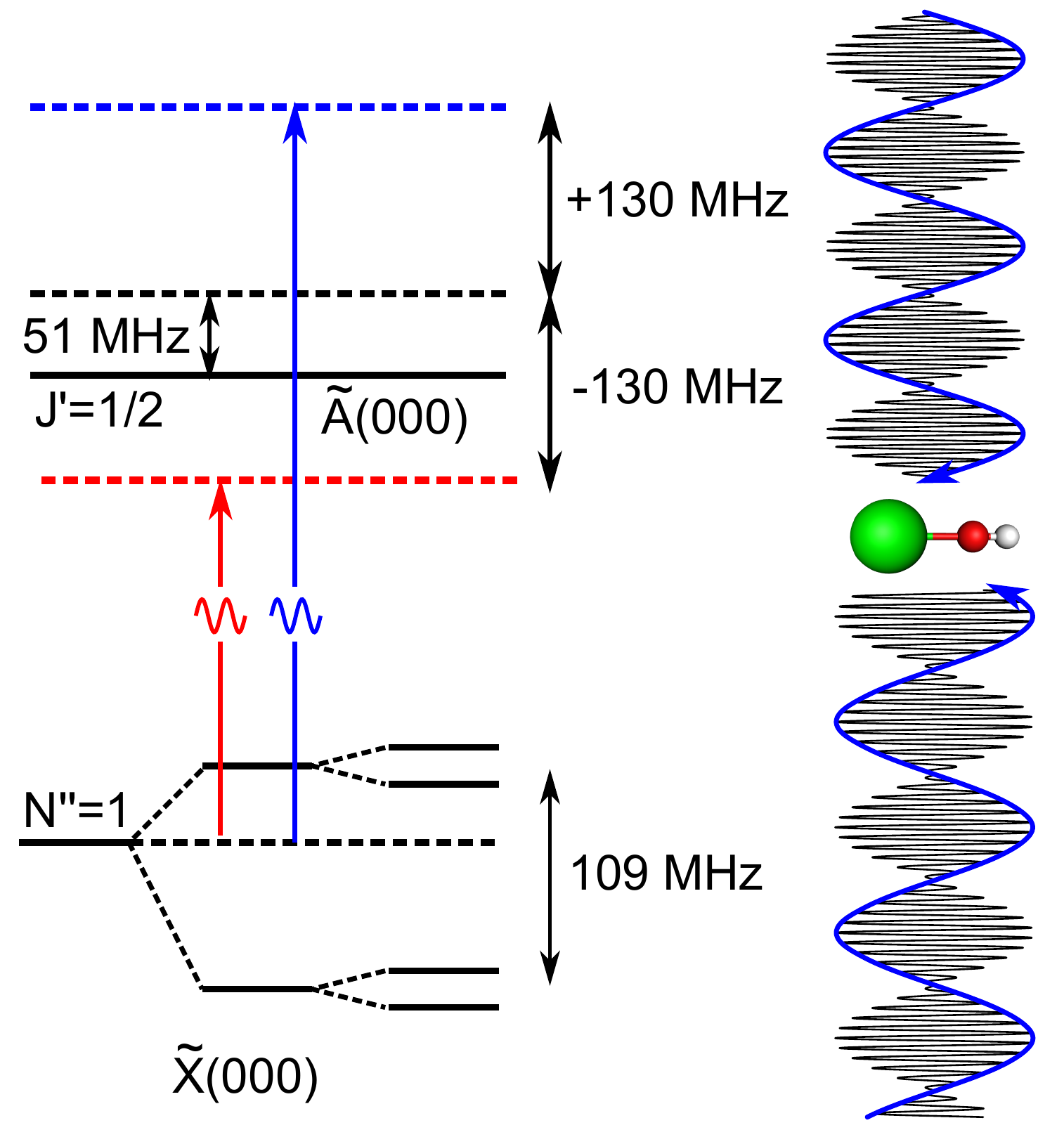}\caption{\label{fig:Detunings_and_beats} The energy diagram on the left-hand side depicts optimal bichromatic frequency detunings for SrOH addressed on the $\tilde{X}^2\Sigma^+\left(000\right) \rightarrow\tilde{A}^2\Pi_{1/2}\left(000\right)$ $P\left(N''=1\right)$ transition for 10 W/cm$^2$ irradiance per each frequency component. The parameters were estimated using numerical solutions of optical Bloch equations. Amplitude modulated optical pulses seen by SrOH molecules are shown on the right-hand side.}
\par\end{centering}
\vspace{-5mm}
\end{figure}

Neutral diatomic molecules are predicted to play an important role in diverse research areas of modern physics such as quantum simulation \cite{bohn2017cold} and computation \cite{demille2002quantum}, as well as searches for new particles and fields beyond the Standard Model \cite{baron2013order}. Larger polyatomic molecules will provide additional opportunities in physics and chemistry \cite{Wall2013,wei2011entanglement,tesch2002quantum}. For example, exploring the origin of biomolecular homochirality \cite{Quack2002} and understanding primordial chemistry leading to the development of organic life requires the use of large molecules \cite{quack1989structure}. However, these molecules' complexity presents significant challenges for direct laser slowing and cooling. Yet these are the key ingredients that allow for optical trapping, which, in turn, realizes long molecule-laser coherence times and high levels of quantum state control. Previously, the external motion of gas-phase polyatomic molecules has been manipulated with off-resonant laser fields \cite{fulton2004optical} as well as electric \cite{Bethlem2000}, magnetic \cite{liu2017magnetic}, and mechanical techniques \cite{chervenkov2014continuous}. Inspired by the success of laser control of atomic motion as well as the latest developments in high-power CW and pulsed laser technology, many experimental proposals and theoretical calculations have been presented on using \textit{stimulated} light forces for molecular beam slowing \cite{chieda2011prospects,dai2015efficient,yang2016bichromatic, aldridge2016simulations, ilinova2015stimulated, jayich2014, romanenko2014cooling}, yet there has been no successful experimental implementation.      

In this Letter, we demonstrate and characterize the optical bichromatic force (BCF) for molecules 
by deflecting a cryogenic buffer-gas beam (CBGB) of SrOH. Using dual-frequency high-power standing light waves we achieve significant force enhancement compared to radiative deflection. The coherent nature of the directional momentum transfer allows multiple $\hbar k$ of momentum change per spontaneous emission cycle.
We also perform theoretical calculations of BCF in complex multilevel
systems and compare to our data. Our results enable accurate predictions for other molecular
species as well as different experimental parameters. Thirty years after the initial theoretical proposal \cite{kazantsev1987rectification}, the bichromatic force for molecules has been conclusively demonstrated.






Radiation pressure beam slowing using white-light \cite{Barry2012,hemmerling2016CaF,Anderegg2017} and chirped techniques \cite{Truppe2017bright, yeo2015rotational} has been achieved for a few simple diatomic species with highly diagonal Franck-Condon factors (FCFs). However, the reduced scattering rate (due to multiple ground state sublevels) leads to a long slowing distance lowering the capture efficiency of molecules from a beam into a magneto-optical trap (MOT) \cite{demille2013transverse}. Significant reduction of the slowing distance in order to mitigate transverse pluming of the molecular beam would result in enhanced MOT densities, desired for diverse applications. Furthermore, expanding beam slowing to species with non-diagonal FCFs (the typical situation for molecules) is extremely challenging due to the rapid loss to dark states after only a few scattered photons.        

\begin{figure*}[t]
\begin{centering}
\includegraphics[width=16cm]{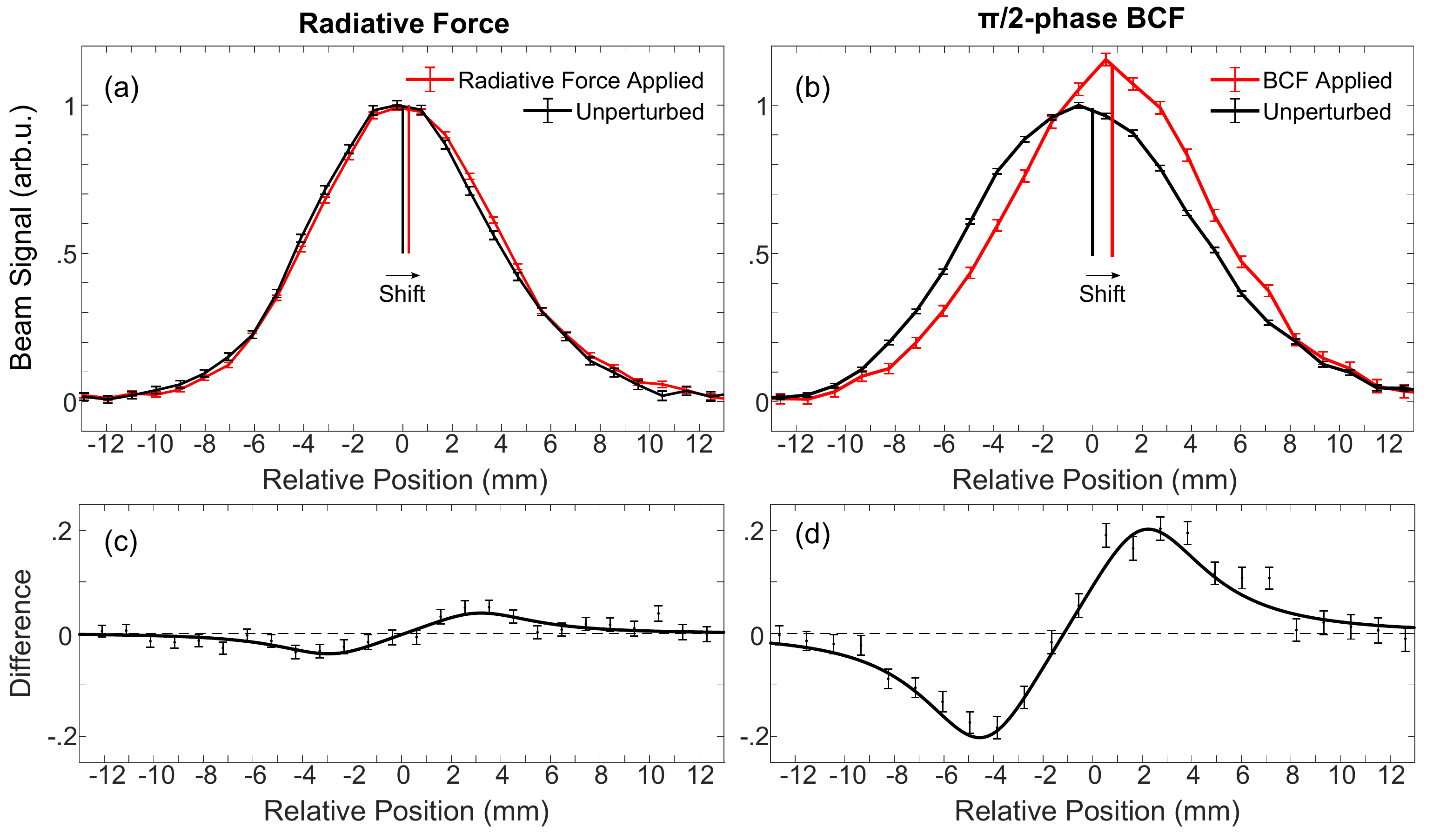} 
\par\end{centering}

\protect\protect\protect\protect\caption{\label{fig:4plots} Data demonstrating bichromatic force deflection of SrOH beam with the $\pi/2$ relative phase shift between the counter-propagating laser beams. Plots (a) and (c) illustrate the effect of purely radiative force on the molecular beam. Plots (b) and (d) compare the beam profiles for unperturbed molecular beam and with BCF applied. The same vertical axis is used for (a) and (b) as well as (c) and (d) plots.}
\end{figure*}

The idea of using \textit{stimulated} slowing and cooling (first for atoms) dates back
to the end of 1980s when the rectified dipole force was theoretically
proposed \cite{kazantsev1989rectification, kazantsev1987rectification} and experimentally observed for sodium atoms \citep{grimm1990observation, voitsekhovich1989observation}.
Using two standing light waves of different frequencies and detunings, the dipole
force can be rectified to apply force on atoms on the macroscopic
spatial scale. The discovery of the bichromatic force (BCF),
which derives from a rapid coherent sequence of absorption-stimulated
emission cycles, led to a technique with a wide velocity capture range and
robustness under experimental imperfections. This allowed significant advances
in atom manipulation \cite{cashen2003optical}. For example, short-distance deceleration of
Cs atomic beam with BCF was demonstrated, achieving a force ten times greater than radiative force  \cite{soding1997short}. Additionally, BCF has been used for He \cite{partlow2004bichromatic} and Ar \cite{feng2017bichromatic} atomic beam collimation.

BCF allows rapid directional momentum
exchange of multiple $\hbar k$ between the light field and atoms
in a single spontaneous emission cycle \citep{soding1997short}. Optical pulses created by the bichromatic fields detuned $\pm \delta$ relative to the transition resonance $\omega$ are shown in Fig. \ref{fig:Detunings_and_beats}. Under the properly chosen laser intensity, each beat envelope of duration $\pi/\delta$ drives a state inversion, transferring $\hbar k$ momenta from the laser field \footnote{While the $\pi$-pulse model presented here is the most intuitive to understand, full description of the bichromatic force on a two-level system involves using a doubly-dressed atom model and can be found in Ref. \cite{yatsenko2004dressed}}. The magnitude of the achieved force becomes $F_{\rm{BCF}}= \hbar k \delta/\pi$ in a two-level system, which can be much larger than $F_{\rm{rad}}= \hbar k \gamma/2$ for $\delta\gg\gamma$. Therefore,
BCF can be effectively applied to molecules that have optically
accessible electronic transitions but suffer from the loss to dark
states (e.g. excited electronic or vibrational levels). Particularly,
application of BCF to polyatomic radicals
that have multiple vibrational modes or diatomic molecules with complex
electronic structures or non-diagonal Franck-Condon factors could allow production of slow, velocity controlled molecular beams
for physics and chemistry applications. 

%
%
%

SrOH provides a convenient platform for exploring the effects of BCF
on complex molecules. As a triatomic molecule, it contains a significant increase in complexity compared to diatomics, including a large number of low-lying bending mode states. A single non-bonding valence electron allows laser addressing
of internal quantum states but the loss to excited stretching and
bending vibrational levels inhibits the photon cycling process \cite{kozyryev2016Sisyphus,Nguyen2017}. 

The exact technical details for the cryogenic production, optical state manipulation, and photon cycling detection of the SrOH beam have been provided in our previous publications \cite{kozyryev2016radiation,kozyryev2016Sisyphus} as well as in the supplementary materials (SM). SrOH in a CBGB is collimated using a $0.5\times2$\,mm aperture before encountering two collinearly superimposed standing waves of 1/$e^2$ Gaussian beam diameter $\diameter\approx0.9$\,mm. We use a bichromatic detuning $\delta=\pm130$\,MHz per each frequency component and a peak on-axis Gaussian beam intensity of 10-11\,W/cm$^2$. The laser light is generated by a tapered amplifier seeded with an injection locked diode laser. An acousto-optic modulator operating at 260\,MHz generates the required frequency offset as described in the SM. Figure \ref{fig:Detunings_and_beats} presents the relevant molecular structure of SrOH and detunings used in the experiment. A magnetic field of $>20$\,Gauss is used to rapidly remix the magnetic dark states created by addressing $J_g>J_e$ molecular structure with a single laser polarization. 

The beat notes interacting with the molecular beam from both directions are generated by the same bichromatic traveling wave upon the reflection from a stationary mirror. The relative phase $\phi$ of the irradiance of the counter propagating traveling waves is controlled by the distance $d_{\phi}$ between the retroreflecting mirror and molecular beam. Since $\phi=4d_{\phi}\delta/c$ \cite{williams2000bichromatic}, the direction of the molecular deflection can be controlled by choosing the appropriate mirror position $d_{\phi}$. Thus, for $\pi/2$ BCF phase shift we use $d_{\phi}=14.4.$\,cm and for $3\pi/2$ phase $d_{\phi}=43.2.$\,cm.

The effect of the BCF on the SrOH molecular beam is illustrated in Fig. \ref{fig:4plots}. The comparison of the radiative force shift (left) with the bichromatic force shift (right) clearly demonstrates larger momentum transfer. This data was taken in the following way. The bichromatic force configuration has two applied beams. To take the radiative force data, the retroreflected beam is simply blocked. Because laser intensities are deliberately balanced, in the absence of the directional momentum transfer, there should be no net shift in the center of the molecular beam. However, we see a deflection of the molecular beam with a larger magnitude than with the radiative force alone.  

\begin{figure}[t]
\begin{centering}
\includegraphics[width=8cm]{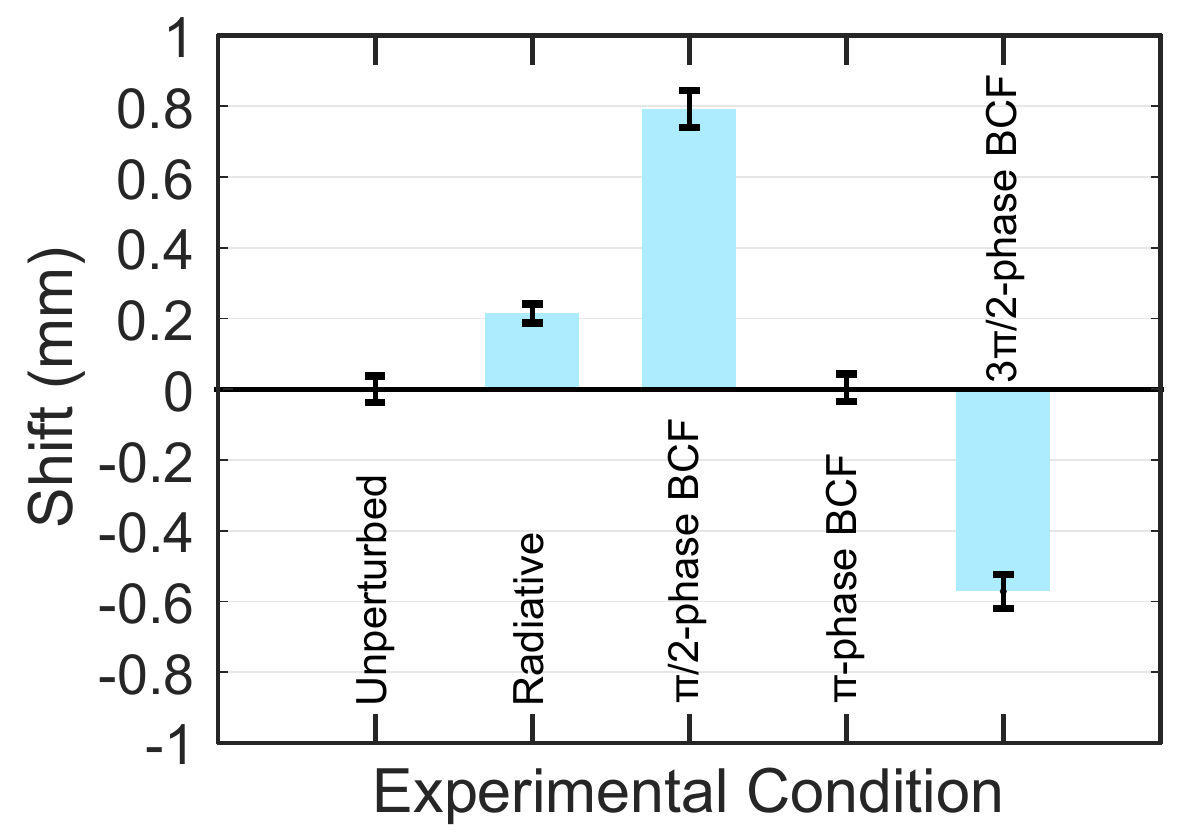}
\par\end{centering}
\protect\caption{\label{fig:Summary_plot} Summary of the molecular beam centroid shift for various experimental conditions: unperturbed molecular beam, radiative deflection with retroreflected beam blocked and BCF deflection with $\pi/2$, $\pi$ and $3\pi/2$ relative phase shift. Reversal of the BCF directionality with the relative phase change is observed.}
\end{figure}

In order to extract the magnitude of the BCF on SrOH and characterize several experimental conditions we plot the absolute value of the beam deflection in Fig. \ref{fig:Summary_plot}. While the direction of the radiative deflection is defined to be positive, by choosing a proper relative phase $\phi=3\pi/2$ between the dual-frequency laser beams transversely addressing the molecules from opposite sides the direction of the BCF can be reversed. The reversal of the deflection direction with $3\pi/2$-phase BCF relative to the radiative force with the addition of the counter propagating beam conclusively indicates the presence of the BCF effect in our experiment. By comparing the ratio of the absolute deflection magnitude for $\pi/2$ and $3\pi/2$ phases we obtain that BCF is $3.7\pm0.7$ and $2.7\pm0.5$ times greater than radiative force, correspondingly. A slightly weaker force for $\phi=3\pi/2$ configuration arises from the larger relative beam diameter imbalance because of the longer $d_{\phi}$ distance. This challenge could be resolved by using separate dual-frequency laser beams shaped identically to the original but with a phase delay introduced. This approach has already been shown to work well for atomic BCF slowing \cite{soding1997short}. For the $\pi$-phase configuration there is no directional momentum transfer between the light beams and molecules and therefore no molecular beam deflection as seen in Fig. \ref{fig:Summary_plot}, as expected for BCF. 

From the BCF deflection data we can estimate that $\left(68\pm5\right)\hbar k$ units of momentum have been coherently transferred from the laser fields to molecules leading to a deflection of 0.3$^{\circ}$. With the clean-up beam for repumping (100) molecules before the camera imaging, we observe no loss of molecules to other dark states within the experimental uncertainties. Thus, for the operating configuration with the repumper present, larger deflection magnitudes can be achieved by extending the molecule-laser interaction time while preserving the proper irradiance condition to achieve efficient stimulated $\pi$-pulse transfer.

The BCF effect on SrOH involves a coherent momentum exchange between molecules and optical fields consisting of two phase-shifted dual-frequency laser beams, which couple 12 ground state magnetic sublevels to 4 excited states. Unlike for atoms \cite{podlecki2017radiation,yatsenko2004dressed}, no analytical solutions exists to describe this process and statistical scaling of the two-level expressions \cite{chieda2011prospects} has proven to be highly inaccurate \cite{aldridge2016simulations}. Therefore, detailed numerical modeling of the bichromatic force in molecules is necessary. In order to estimate the bichromatic force effect on SrOH we have employed direct numerical solution for the time-dependent density matrix in the rotating-wave approximation \footnote{Ref. \cite{aldridge2016simulations,aldridge2016bichromatic} and SM provide the details of how the calculations for SrOH were performed}.

\begin{figure}
\begin{centering}
\includegraphics[width=0.4\textwidth]{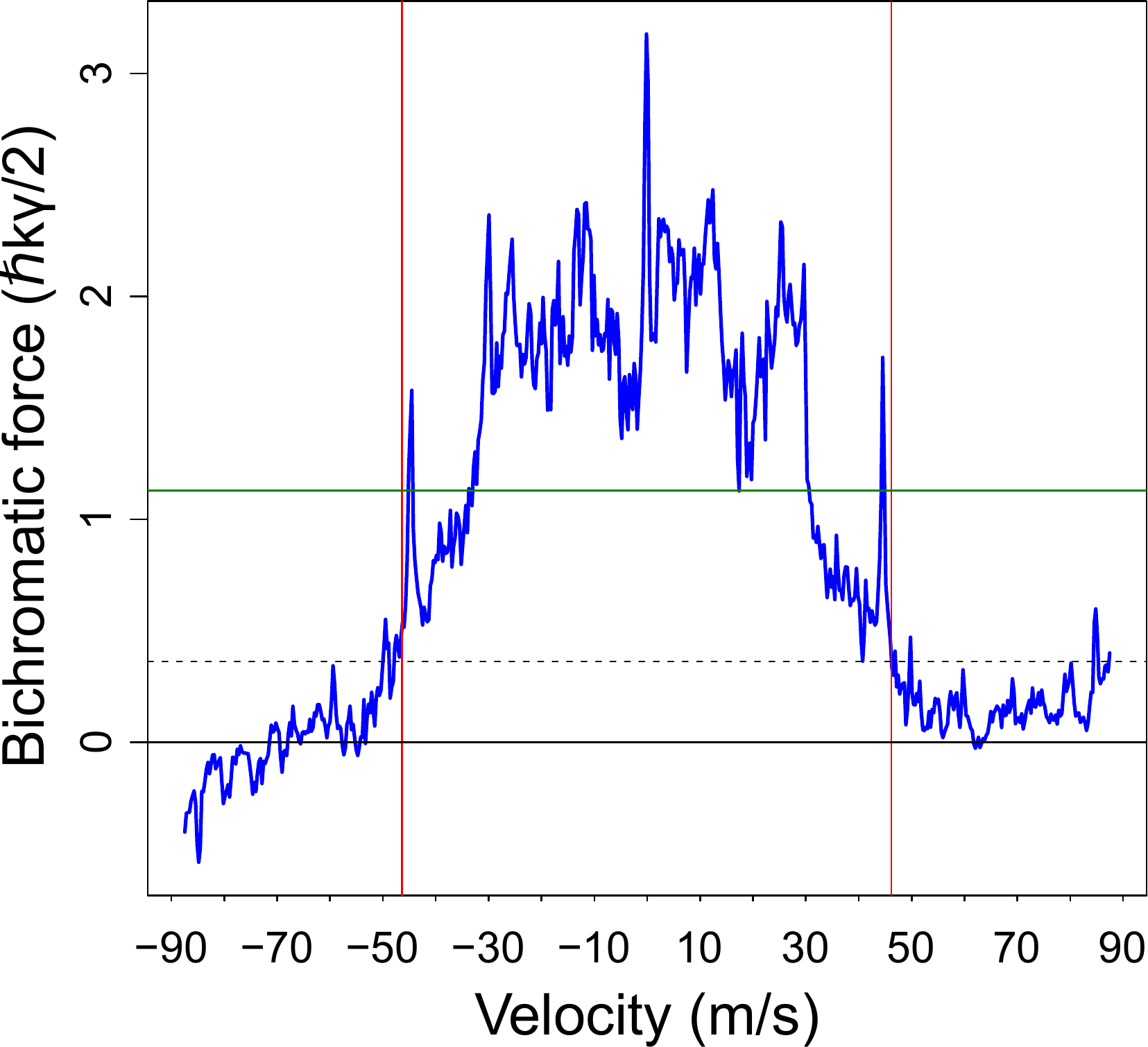}\protect\caption{\label{fig:BCF_vs_velocity} Calculated profile of the bichromatic force exerted on SrOH molecules as a function of velocity under optimal experimental conditions and achieved laser power. The dashed horizonal line indicates the magnitude of the maximal radiative force for SrOH in our experiment. Two vertical red lines mark the region where the magnitude of the BCF exceeds purely radiative pressure. Sharp resonances around 0 and $\pm45$ m/s represent Doppleron resonances and have been previously seen in BCF profiles for atoms \cite{feng2017bichromatic}. Solid horizontal green line shows the average bichromatic force achieved in the experiment. The calculated force profile assumes an irradiance of 10 W/cm$^2$, a magnetic field of 27.2 G skewed 33$^{\circ}$ with respect to the polarization axis and the frequency parameters depicted in Fig. 1. }
\par\end{centering}
\end{figure}

Figure \ref{fig:BCF_vs_velocity} presents the results of the numerical calculations for the BCF effect on SrOH as a function of molecular velocity along the laser propagation direction. The average BCF force exerted on molecules with velocities $v\leq25$ m/s is $\left(1.89\pm0.04\right)\hbar k \gamma/2$, which exceeds the maximum radiative force in the experiment by 5.7 times. Using the measured optimal BCF deflections for SrOH we obtain $F_{\rm{BCF}}=\left(3.7\pm0.7\right)F_{\rm{rad}}$, which is 35$\%$ lower compared to the predicted value. Experimental observations of BCF for atoms have seen significant reduction in the force magnitude under realistic experimental conditions \cite{feng2017bichromatic}. In our experiment, some of possible explanations for discrepency are power imbalance between the two beat note trains, less than unity overlap efficiency between the laser and molecular beams, and a dependence of the two forces on irradiance which favors the radiative force in the wings of the Gaussian laser profile. 

Our experimental measurements and theoretical calculations indicate compression of the transverse velocity distribution of the deflected SrOH beam. From Fig. \ref{fig:4plots} we conclude that the full width at half maximum of the beam distribution is reduced from 10.6$\pm$0.2 mm to $9.3\pm$0.2 mm. In earlier experiments, both longitudinal \cite{soding1997short} and transverse \cite{partlow2004bichromatic} cooling of atomic beams have been previously observed, including in the ultimate limit of $\lesssim$1 scattered photon \cite{corder2015laser}. In combination with previous atomic experiments \cite{partlow2004bichromatic}, our results indicate the possibility of creating optical molasses-like bichromatic field configurations characterized by rapid damping forces and wide velocity capture range using two sequential spatially separated regions with opposite $\phi$.    

In summary, we demonstrate a coherent optical force on polyatomic molecules using bichromatic laser fields. SrOH has 12 ground state sublevels coupled to 4 excited states, yet we achieve $F_{\rm{BCF}}\approx1.1\hbar k \gamma/2$ which is effectively the size of the radiative force on an ideal two-level system. This can be compared with other radiative force experiments on diatomic  \cite{Shuman2009}  and polyatomic molecules \cite{kozyryev2016radiation}, which have previously shown $F_{\rm{rad}}\approx\hbar k \gamma/7$ and $\hbar k \gamma/4$, correspondingly. Thus, large resonant optical forces on molecules are possible with the use of coherent bichromatic optical fields to induce stimulated forces. 

The magnitude of the bichromatic force achieved in this experiment is  limited by the available optical power from our laser system. Our calculations indicate that with 1\,W per each frequency component and using a 2\,mm 1/$e^2$ Gaussian beam diameter, the average value of the BCF will be $7\,\hbar k\gamma/2$ for molecules spanning a 60 m/s range \footnote{See Ref. \cite{aldridge2016bichromatic} and SM.}. Slowing of a cryogenic beam of SrOH from 60 m/s to near zero in a distance of $\lesssim 5$\,mm could be achieved with only 500 scattered photons, compared to 10,000 photons (and a much longer distance) required for radiative slowing. The use of polyachromatic forces with higher harmonics could further enhance the magnitude of stimulated forces by reducing the overlap of counter-propagating beat pulses, in addition to extending the velocity capture range \cite{galica2013four,aldridge2016bichromatic}. The powerful combination of cryogenic buffer-gas beam technology and stimulated optical forces has the potential to deliver slow molecules with complex internal structures for precision beam measurements or for trapping,  e.g. for slowing heavy polyatomic molecules like YbOH to search for new fundamental particles at the PeV energy scale \cite{kozyryev2017precision}.    

\begin{acknowledgments}

This paper is dedicated to Professor Eyler, who astutely recognized the immense merit of stimulated forces for slowing molecular beams and enthusiastically propelled this field forward on both experimental and theoretical frontiers. We would like to thank Kyle Matsuda and Yicheng Bao for building the TA laser system. We also thank Scott Galica, Zack Lasner, Loic Anderegg, and Ben Augenbraun for many insightful discussions. This work has been supported by the NSF, grant \# PHY-1505961. 

\end{acknowledgments}

\bibliographystyle{apsrev}
\bibliography{SrOH_BCF_library}

\end{document}